\documentclass[sigconf]{acmart}

\usepackage{xspace}
\usepackage{cleveref}
\usepackage{listings}
\usepackage{subcaption}
\usepackage{soul}
\usepackage{mdframed} 

\usepackage[table]{xcolor}
\usepackage{booktabs}
\usepackage{siunitx}
\usepackage[most]{tcolorbox}
\usepackage{soul} 
\usepackage[normalem]{ulem}
\usepackage{textcomp}

\usepackage{float}
\usepackage{tabularx}

\usepackage{mdframed}

\usepackage{menukeys}

\newlength\MAX  

\newcommand{\mysec}[1]{\vspace{0.08cm} \noindent \textbf{#1.}}

\newcommand{\tool}{Motif\xspace}

\usepackage{wrapfig}

\usepackage{ifthen}
\newboolean{showcomments}
\setboolean{showcomments}{true} 
\ifthenelse{\boolean{showcomments}}
  {\newcommand{\nb}[2]{
    \fcolorbox{gray}{yellow}{\bfseries\sffamily\scriptsize#1}
    {\sf\small$\blacktriangleright$\textit{#2}$\blacktriangleleft$}
  }
  
  }
  {\newcommand{\nb}[2]{}
  
  }

\newcommand{\ie}{i.e.,\xspace}

\definecolor{added}{HTML}{AAFFAA}
\definecolor{deleted}{HTML}{FFAAAA}
\definecolor{edited}{HTML}{FFDDB3}

\lstdefinestyle{normal}{
  language=python,
  basicstyle=\ttfamily\scriptsize,
  aboveskip=0pt,
  belowskip=0pt,
  escapeinside={(*}{*)},
}

\lstdefinestyle{a}{
  language=python,
  basicstyle=\ttfamily\scriptsize,
  aboveskip=0pt,
  belowskip=0pt,
  backgroundcolor=\color{added},
  escapeinside={(*}{*)},
}

\newtcbtheorem{Summary}{\bfseries Summary}{enhanced,drop shadow={black!50!white},
  coltitle=black,
  top=0.15in,
  attach boxed title to top left=
  {xshift=1.5em,yshift=-\tcboxedtitleheight/2},
  boxed title style={size=small,colback=white}
}{summary}

\newtcolorbox[auto counter]{prompt}[1][]{title={\bfseries},enhanced,drop shadow={black!50!white},
  coltitle=black,
  top=0.1in,
  attach boxed title to top left=
  {xshift=1.5em,yshift=-\tcboxedtitleheight/2},
  boxed title style={size=small,colback=white},}

\usepackage[normalem]{ulem}

\usepackage{tikz}

\definecolor{ABlue}{HTML}{127bca}
\definecolor{LHScolor}{HTML}{555555}

\usepackage[scaled]{helvet}
\usepackage[most]{tcolorbox}

\setcopyright{none}
\renewcommand\footnotetextcopyrightpermission[1]{} 
\begin{document}

\title{Motif: Discovering and Automating Personal Web Workflows}
\author{Shaokang Jiang}
\email{shj@uci.edu}
\affiliation{%
  \institution{University of California, Irvine}
  \city{Irvine}
  \state{California}
  \country{USA}
}

\author{Daye Nam}
\email{daye.nam@uci.edu}
\affiliation{%
  \institution{University of California, Irvine}
  \city{Irvine}
  \state{California}
  \country{USA}
}

\begin{abstract}
  Recent advances in LLMs and existing work on programming by demonstration have made it possible for end users to create automations by explicitly demonstrating their behavior to LLMs. However, these approaches rely on the assumption that users know what to automate and what is capable of being automated. Additionally, automation via LLM agents is often expensive compared with programs.
  We introduce Motif, a system that passively observes everyday browser activity to discover recurring interaction patterns that are programmable, makes recommendations to users whenever a pattern is discovered and generate a program to install after user confirmation. Users can review, and refine the program using natural language. We evaluated Motif in a multi-day study, comparing its ambient discoveries against automations users attempted to build via ``vibe coding.'' With eight participants, Motif discovered more automatable patterns than users recognized. Most of them matched participants’ routines and were useful. Follow-up surveys showed most would continue using Motif-generated programs.

\end{abstract}

\begin{teaserfigure}
    \centering
  \includegraphics[width=0.9\textwidth]{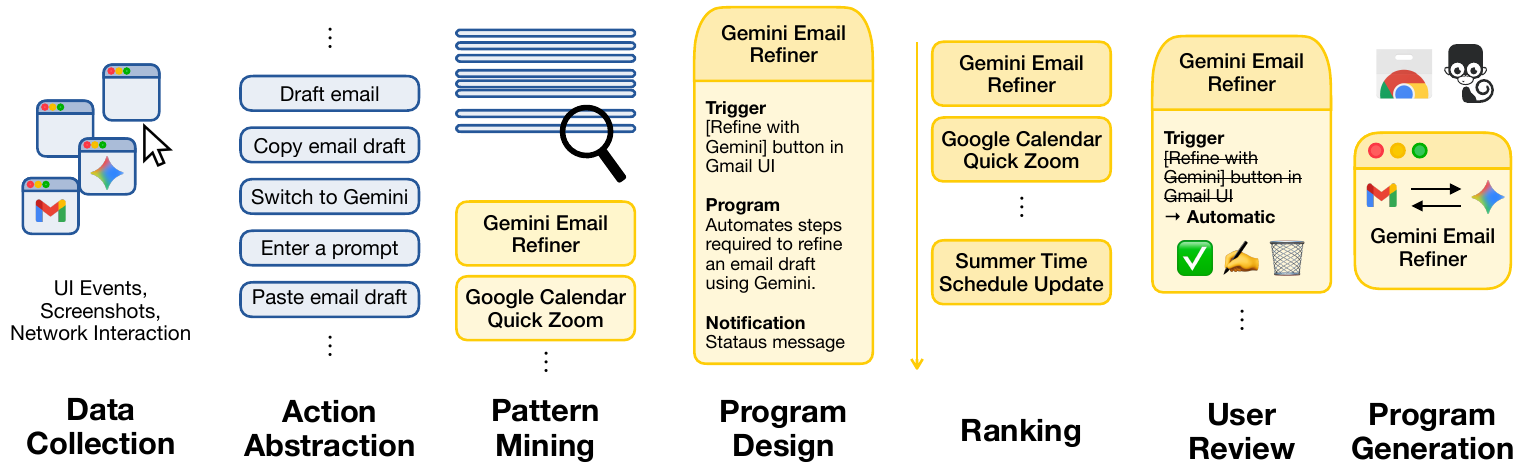}
  \caption{Overview of \tool. \tool supports ambient automation discovery by streaming a pipeline of passive data collection, LLM-based pattern mining, and program generation. The interface lets users review, refine, and deploy discovered automations.}
  \label{fig:teaser}
\end{teaserfigure}

\maketitle

\section{Introduction}

The rise of LLM (Large Language Model)-powered code generation is making programming more accessible. 
``Vibe coding~\cite{Liu_2023,horvat2025vibe}'', promises that end users may soon be able to describe what they want in natural language, and receive working code in return, using LLM-powered tools like GitHub Copilot, Cursor, Claude Code, Antigravity, and many more.
These tools have demonstrated that natural language could function as a way of programming, holding particular promise for end-user programming.

However, vibe coding retains a critical assumption from traditional software engineering: the user must know \textit{what} to build first~\cite{ko2004six, pea1987user, Liu_2023, shome2025johnny}.
For example, a business analyst must recognize that their daily routine of checking three dashboards and compiling a summary email is a candidate for automation. 
Or, a real estate agent must recognize that their workflow of copying property details into a comparison spreadsheet from multiple listing services and pasting the batches into an email could be a script, rather than feeling like ``doing my job.''
Some of these tasks are so embedded in daily practice that they feel like work itself rather than overhead that could be eliminated, and even developers are not immune.
Furthermore, for end users without programming experience, these patterns are invisible because they lack a mental model of what a program can do.

Programming by demonstration (PBD) systems have sought to make automation accessible to end users~\cite{li2017sugilite}.
However, they also require the user to initiate the process, and users must first be aware that such tools exist, then recognize a task as worth recording, and then initiate the recording.
Whether through vibe coding or PBD, the bottleneck is not \textit{how} to program, but knowing \textit{what} to program in the first place.

In this paper, we propose a new way of end-user programming, through \textbf{ambient automation discovery}; rather than requiring users to identify and describe what to program, we build a system that passively observes user actions, automatically discovers recurring multistep patterns, and surfaces them as candidate programs.
The system automatically identifies what is programmable and ``program-worthy'' so users do not have to.
By converting these patterns into deterministic programs, the discovered automations become artifacts that can be deployed, revisited, and shared with others~\cite{leshed2008coscripter}. Moreover, because all programs are executed locally, they can reduce privacy risks~\cite{kleppmann2019local} and be more cost-efficient than LLM agents that incur inference costs with each execution~\cite{chen2023frugalgpt, li2025alloy, liu2025reuseit}.

This paradigm of including both textual requirement and behavior incidence produces better programs even when people know what they want.
For example, when a user wants a tab manager, the user would likely under-specify their needs, as describing their routine is a big challenge.
The user would describe their routine as ``I want to clean up my tabs sometimes'' rather than articulating the precise six-step sequence they follow across specific sites.
The users may be able to build a generic tab manager through vibe coding, but it will not be grounded in what they actually did, and they will need to spend a long time refining it as shown in previous work~\cite{Liu_2023, shome2025johnny, 10.1145/3586183.3606822}.
Under this paradigm, because the discovered patterns are grounded in the user's actual behavior, the resulting specifications are inherently personalized.

To realize this vision, we present \tool, a system that passively records everyday browsing activity, uses LLMs to identify recurring automatable patterns, design user-friendly programs, and provides a Chrome extension frontend where users can review, refine, and deploy them.

We evaluated \tool in a study with eight participants over an average of 5.5 (M = 5, SD = 1.2) days of real browser usage.
In total, the system identified 175 patterns, averaging 22 per person. 
During the user study, participants went through a total of 40 patterns. 
Among the 40 patterns, the participants agreed that 34 (85\%) corresponded to their daily routine. 
Participants tried to build programs for these 40 patterns, and 24 (60\%) were successfully deployed and run as expected. 
Notably, participants were able to recognize only three (7.5\%) of the 40 patterns as automatable on their own before using our tool, confirming the what-to-build problem. A follow-up survey revealed that most of the programs remain functional and will continue to be used.

In summary, this paper makes the following contributions:
\begin{itemize}
    \item The concept of \textit{ambient automation discovery} from end-user workflows, addressing the initiation gap between programming by demonstration and end-user programming.
    \item A technical pipeline that combines textual requirements with user-grounded behavioral evidence mined from observed actions, enabling LLM-based code generation to produce better personalized, deterministic, and deployable programs.
    \item An interaction design for reviewing, previewing, and iteratively updating discovered programs.
    \item An empirical evaluation with eight participants demonstrating that the system discovers meaningful, automatable patterns that users would not have identified on their own.
\end{itemize}

\section{Backgrounds and Related Work}

\subsection{End User Programming}
End-user programming, enabling people without professional software development training to create programs that meet their own needs, has been a long-standing goal in HCI.
Over the past decades, significant progress has been made in lowering the barriers to program creation.
One noteworthy example is Programming by Demonstration (PBD), where users specify their needs by showing the system what to do~\cite{li2017sugilite, li2019pumice, herskovitz2024programally, 10.1145/3586183.3606822}.
Trigger-action programming further simplifies creation by decomposing automations into if-this-then-that rules~\cite{aveni2025generative, trigger1, trigger2, trigger3}.
More recently, the LLM-powered code generation tools allow users to describe programs in natural language and receive working code in return, or a practice increasingly referred to as ``vibe coding.''~\cite{horvat2025vibe}
These approaches have lowered the barrier to \textit{creating} programs for end users.

\subsection{Initiation Gap}
However, these systems share a common upstream assumption: the user must first recognize that a task is worth automating and then initiate the process.
This assumption, however, is often far from trivial, due to several compounding factors that form what we call the \textit{initiation gap}:
\textbf{1) Task habituation}: Repetitive tasks become invisible through familiarity. When a task is performed frequently enough, it can feel like ``just doing work'' rather than overhead that could be eliminated, even when they require significant time or effort when aggregated. Research on habituation suggests that people systematically fail to notice their own repeated patterns, and self-tracking studies consistently show that users are surprised by their own behavioral data~\cite{10.1145/2533670.2533672, 10.1145/1753326.1753409}. \textbf{2) Mental model gap}: End users without programming experience lack a mental model of what a program can do. Without this mental model, end users cannot recognize which of their workflows could be handled by a program, even when those are clear to a programmer~\cite{pea1987user, Liu_2023, ko2004six}.
\textbf{3) Specification gap}: Even when a user can recognize what to program, articulating it precisely enough can be difficult for end users. Studies on prompt engineering show that non-programmers struggle to specify their needs in natural language~\cite{shome2025johnny, zamfirescu2023johnny, jiang2025beyond}, and even studies on PBD tools report the challenges of end users in demonstrating their needs in a generalizable manner~\cite{li2017sugilite, 10.1145/3242587.3242661}.
\textbf{4) Automation overhead}: Even when users overcome the above barriers, users, even programmers, sometimes rationally avoid initiating automation: the upfront cost of setting up tools for end-user programming, or writing programs, can feel higher than simply performing the task one more time~\cite{blackwell2002first, mackay1991triggers}, even when the cumulative cost is significant~\cite{hellendoorn2019code,  lau2009programming, jiang2024analysis}. 

Together, these factors constitute the \textit{what-to-automate} problem. 
Although existing tools have solved \textit{how} to create programs to automate workflows (\ie automation)~\cite{li2025alloy, liu2025reuseit, huq2026modelingdistincthumaninteraction}, and commercial product have enable end-user to generate replable script~\cite{browserflow2025, axiom2025}, it remains open to identify which tasks are candidates for which to program and when to actually initiate the programming process.

\subsection{Usage Pattern Mining}

If users cannot identify what to automate, can systems do it for them?
For well-defined business processes, this idea has been used in Robotic Process Automation (RPA) discovery~\cite{leno2021robotic}.
These systems actively look for ``what can be automated?'' by extracting workflows from event logs.
However, these approaches assume well-defined business processes with clean event schemas, typically evaluated and operated by domain experts~\cite{syed2020robotic, agostinelli2022reactive, le2023log}, which is a fundamentally different setting from everyday end-user automation.

Previous work on usage mining and clickstream analysis has studied extracting user behavior patterns from logs~\cite{10.1145/1502650.1502690, choi2021candidate, bosco2019discovering}, pre-collected interaction traces within a single app~\cite{huang2024automatic, arsan2021app}. These techniques have been applied to product evaluation~\cite{kohavi2009controlled}, user modeling for personalization~\cite{tokucc2025predicting, 10.1145/345124.345169} or UX optimization~\cite{10.1145/1135777.1135811, turbeville2024llm}, semantic interpretation of user actions~\cite{10.1145/1111449.1111492, 10.1145/3706598.3714213},  behavior prediction~\cite{10.1145/1111449.1111473}, and validation of the semantic meaning of extracted patterns~\cite{huang2024automatic}.

However, to the best of our knowledge, prior work has not fully addressed the complexity of real-world user interactions or explicitly designed approaches for surfacing automation opportunities from end users' daily behavior. Existing approaches typically require users to possess sufficient background knowledge to recognize potential automation opportunities and to initiate data collection themselves~\cite{huang2024automatic, 10.1145/3242587.3242661}. In contrast to those post-hoc recognition, which often depends on user initiation and a well-curated corpus, we focus on discovering recurring patterns from real end users' daily behavior across multiple applications and surfacing them as candidate programs for automation.

One notable system is LiveAction~\cite{10.1145/2533670.2533672}, which introduced a fully automated approach for generating task models directly from real-world web usage logs. While it relied on classical sequence mining to model user behavior step-by-step, the advent of Large Language Models (LLMs) could enable a much richer, semantic approach to pattern extraction~\cite{akhtar2025llm}.

\section{Ambient Automation Discovery for End Users}

As discussed in the previous section, existing approaches to end-user automation assume that the user must recognize a task is automatable and initiate the process~\cite{li2017sugilite, li2019pumice, huang2024automatic}.
We believe that this gap prevents broader adoption of end-user programming or automation.

To address this, we propose \textit{Ambient Automation Discovery}: a paradigm in which the system observes everyday user behavior, identifies recurring patterns that are candidates for automation, and deploy them as programs after users review and confirmation.
We derive four design goals for this paradigm from the gaps we identified in our literature review:

\begin{itemize}
    \item \textbf{DG1: Discover without user initiation.} The system should identify automatable patterns without requiring users to recognize or declare them. 
    \item \textbf{DG2: Generate personalized specifications from observed behavior.} The system should produce program based on what users do, not generic templates.
    \item \textbf{DG3: Keep users in control.} Despite passive observation, users should be able to review, understand, edit, and reject discovered automations before deployment.
    \item \textbf{DG4: Produce deterministic and cost-efficient programs.} The output should be a persistent artifact that users can revisit, modify over time, stable, and low-cost.
\end{itemize}

\subsection{Usage Scenario}
Consider Alex, a graduate teaching assistant who regularly receives emails from students requesting accommodations for missed quizzes. 
For each request, Alex drafts a reply in Gmail, then opens Google Gemini in a new tab, copies the draft, pastes it with a prompt asking Gemini to make the reply more professional and empathetic, waits for the result, copies the refined text, and pastes it back into Gmail. 
Alex does this several times a week, but has never thought of it as
something a program could handle, as it simply feels like part
of replying to students.

After a few days of normal browsing with \tool installed, Alex opens the \tool interface and sees a ranked list of discovered programs. 
Near the top is one titled ``Student Accommodation
Email Refiner.'' 
The system has identified the recurring pattern, drafting a reply, switching to Gemini, and copy-pasting text back and forth, specifically in the context of accommodation request threads. 
Alex is surprised, because they never considered this a ``task'' that could be automated.

Alex opens and reviews the program design generated by \tool: 
(Figure~\ref{fig:teaser}-Program Design):
a ``Refine with Gemini'' button injected into Gmail's compose toolbar that, when clicked, automatically copies the current email draft to the Gemini web interface and transfers the polished version back.
The trigger, program behavior, and notification are all
described in natural language. 
The description is specific to accommodation emails, but Alex realizes this would be useful for \textit{all} email drafts, not just accommodation replies.
Alex edits the program design to remove the accommodation-specific language and generate the program.
After confirming the updated design, Alex clicks ``Generate Program,'' installs the script, and from that point on, a single click in Gmail around the original send button replaces the multi-step, multi-tab routine.

In this scenario, Alex never had to recognize this workflow as automatable, describe it from scratch, or initiate any recording. 
\tool surfaced the opportunity from observed behavior (\textbf{DG1}), generated a specification grounded in what Alex actually did (\textbf{DG2}), and let Alex review and refine it before deployment (\textbf{DG3}), producing a deterministic script that runs locally at no ongoing cost (\textbf{DG4}). 
Notably, the observation-grounded starting point, a specific pattern tied to accommodation emails, gave Alex a concrete foundation to generalize from, rather than having to articulate a vague request like ``help me polish my emails'' to generate and reiterate the program from scratch, nor needing to specify personal preferences, such as they prefer Gemini over ChatGPT.

\section{\tool: Ambient Automation Discovery}
\tool transforms raw browser activity into deployable automation programs through a pipeline described in \Cref{fig:teaser}.

 \begin{figure*}[t]
\includegraphics[width=0.9\linewidth]{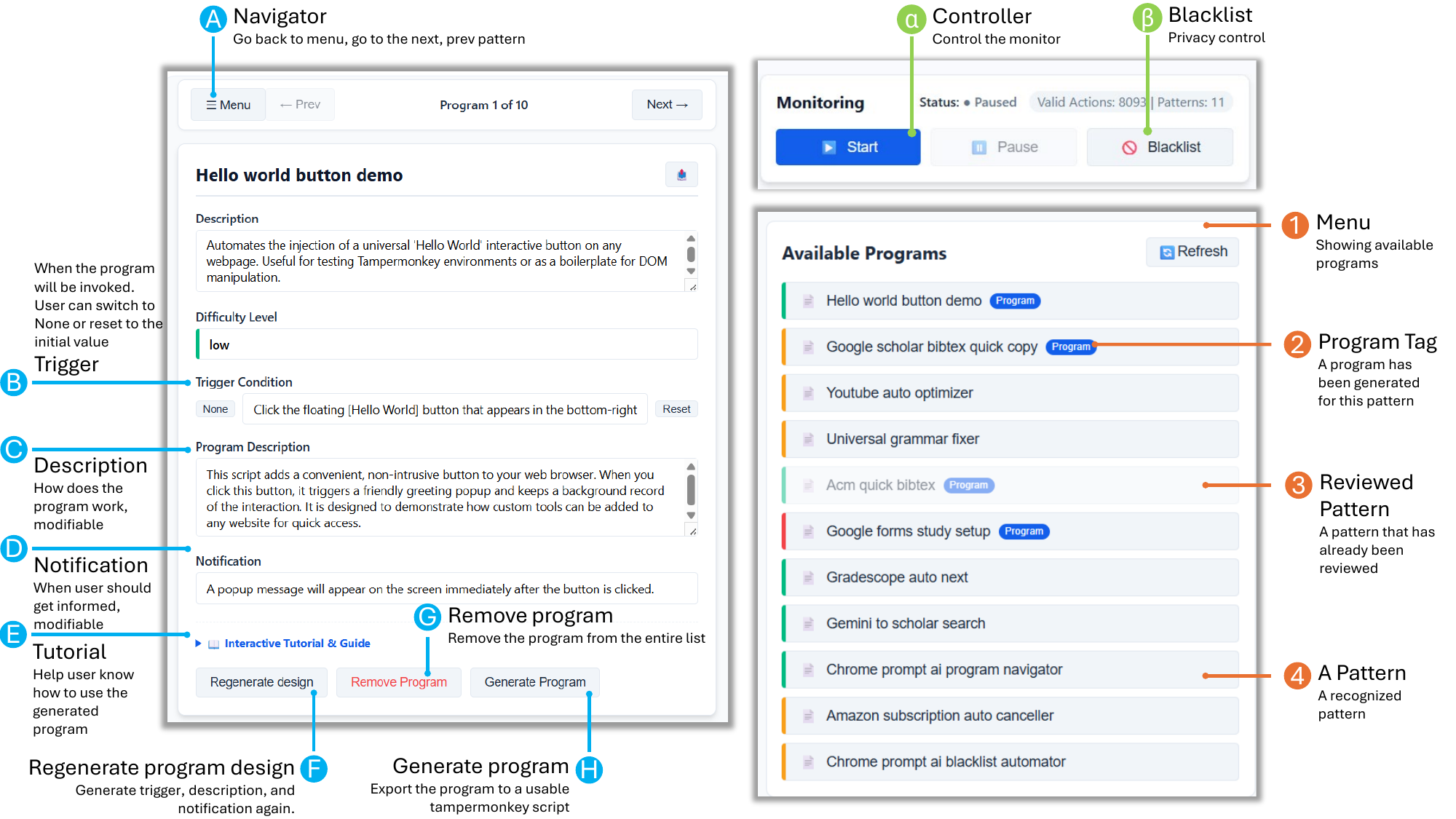}
\caption{User interface. Right top ($\alpha-\beta$): controls for managing system monitoring and data collection. Right bottom (1-4): menu for reviewing generated programs. Left (A-H): an interface showing discovered programs once the user clicks on a pattern. }
\label{fig:interface}
\end{figure*}

\subsection{Data collection}

To capture the web usage with enough context, \tool collects four sets of logs.
First is UI interactions, such as clicks and copy-pastes.
When extracting these interactions, \tool also captures surrounding text and corresponding IDs or ARIA labels to provide sufficient context. 
Second is browser system interactions, such as switching or opening tabs, along with the URLs.
\tool also collects OS-level interactions, including screenshots of the Chrome window with absolute mouse position, the time when user leaves the browser, refocuses the browser, etc. 
As a backbone of the information, \tool also collects lite network interactions, including API requests and responses, with corresponding URLs, request bodies, response bodies, etc., within 1.5 seconds after the UI interactions.

\subsection{Action Abstraction}
Due to the complexity of user interactions in each chunk, \tool abstracts low-level interactions into high-level actions.
\tool leverages LLM multimodal capabilities as the primary method for understanding user interactions and summarizes detailed interaction logs into a natural-language description with metadata to prevent hallucination, such as ``clicking the 'Post' button on Ed Discussion, with metadata such as copied text content and clicked page URLs.''
\tool builds an action database containing a sequence of user actions with metadata.

\subsection{Pattern Mining}
\tool then uses LLMs to identify all programmable sequences as patterns. 
The LLM is instructed to focus on automatable sequences, find and merge newly added programming sequences to the previous similar patterns.
This approach differs from traditional sequence mining algorithms, which rely purely on sequence frequency. 
While traditional sequence mining algorithms could be an alternative, it is easier to filter out noise and 
identify more meaningful and deterministic patterns that are programmable using LLM, leveraging programming as a reasoning mechanism. 
Throughout this step, \tool builds a pattern database 
containing all identified patterns with metadata such as each individual occurrence.

\subsection{Program design}

Patterns obtained from the previous step are usually descriptions of sequences of user interactions, which are not directly programmable, as they are just plain English descriptions of actions users performed. 
For patterns that are likely to be frequent, we leverage LLMs as a designer to analyze the identified patterns and generate user-friendly program designs without technical details while considering the technical feasibility, which include the following components:

\mysec{Trigger}
A button or a shortcut. 
For some tasks, it can be automatic. 

\mysec{Program Description} 
It provides detailed and non-technical descriptions that are necessary for program generation.

\mysec{Notification} It shows the situation that user is going to be notified, such as successfully completed. Given that some programs are automatically inferred without actions, it is important to notify users when any automation occurs.

\subsection{Program Ranking}
Given that many patterns can be discovered through passive observation and that users typically review only the top selections~\cite{pernice2014people}, \tool displays the candidate programs sorted by a score determined by the LLM based on a combination of frequency and the likelihood of future occurrence.

\subsection{User Review}
To ensure we give users control, \tool only generates programs after the user confirms the program design.
More details in supporting user review can be found in \Cref{subsec:ui}.

\subsection{Program Generation} 
Once the user confirms the program design, \tool sends a request to an LLM with the program design and context collected in the previous steps, to generate a program to automate the pattern.
Among many approaches that can achieve automation, including agent workflow as in \citep{li2025alloy}, we chose to implement automation programs to achieve high cost-efficiency and reusability (DG4).
Programs are sustainable at scale, run faster, privacy-guarded, and are significantly less costly compared to agent workflows, which require agent API usage and incur charges per execution.

\subsection{User Interface} 
\label{subsec:ui}
To ensure users still have control over the ambient discovery process, as well as what to automate, we carefully designed a front-end interface for \tool (\Cref{fig:interface}).

\mysec{System controller ($\alpha-\beta$)}
To address privacy concerns, we give users control during the monitoring process. Users can pause monitoring at any time and add URLs to a blacklist to prevent the system from monitoring their interactions on those pages. A status indicator is also displayed to show whether the system is currently monitoring or paused, together with information such as the number of patterns discovered.

\mysec{Menu for reviewing generated patterns (1-4)}
Once discovered patterns are available, the menu displays an entry for each pattern, presenting a list of ``Available Programs.'' Each entry contains a descriptive title, a color-coded difficulty indicator on the left, and a blue ``Program'' tag if there is a program associated with the pattern generated earlier. 
Patterns that were reviewed before appear faded.

\mysec{Program design review and refinement (A-H)}
When users click on a pattern entry, a detailed interface will be displayed, showing the program design generated by the LLM.
Users can modify all aspects of the program design, 
as well as to request a new design from the LLM, which will generate a brand-new program design that still meets the logic. 
Once the user is satisfied with all descriptions, they can deploy it by clicking "Generate Program."

\mysec{Privacy and Security}
\label{subsec:privacy}
We designed Motif as a research prototype to validate ambient discovery rather than a production-ready system. As a result, we implemented only a minimal set of privacy safeguards, including filtering passwords and hidden fields before recording, truncating network request bodies and URLs, discarding responses prior to LLM calls, requiring human validation of designs before program generation, and asking participants to avoid sensitive content while running the study. Each generated Tampermonkey script also includes several LLM-determined \texttt{\@grant} scopes, which ensure the script only has access to the scopes for which it is intended. All data are stored locally on the user's machine, and users can inspect, review, revise, and delete data at any time.

\subsection{Implementation}

\tool is implemented as a Chrome browser extension with a cross-platform native host, 
given the limited, inefficient storage available in Chrome extensions, the limited screenshot capabilities of Chrome extensions.
The frontend was built with JavaScript, HTML, and CSS. 
To ensure good cross-platform compatibility, the native host was built with Rust and communicates with the frontend through Chrome Native Messaging. Users can inspect, review, and manage all data stored locally.

For most of the pipeline, when LLM was involved,
Gemini 3 Flash Preview was used with high-thinking enabled.

For action abstraction, all raw events are unified with a single Unix timestamp and grouped into several small chunks based on a maximum of 60 seconds of continuous activity, whenever a user switches away from and returns to the browser, 15 seconds of inactivity, or 8 consecutive screenshots. 
Then, each chunk was sent to LLM for abstraction, with the maximum context size to be under 3K tokens.
Small context windows also allow future deployment to consider local LLM deployment, which usually has more limited context windows, more sustainable, and cheaper.

For pattern identification, we enabled tool usage of code interpreter, URL scraping, and Google Search to help the LLM better understand user interactions and find more accurate patterns. 
Whenever 500 new actions are discovered, we send a single request to the LLM to perform pattern identification. 500 was selected after several iterations. When generating responses for the previous two steps, we used a temperature of 0.1 to enable stable and accurate abstraction. 

To ensure full streaming and asynchronous processing, \tool was set to automatically check for new data using the above two procedures every 5 minutes and update the relevant database accordingly. 
All uploaded images were cleaned right after finishing those two procedures to prevent caching and protect user privacy.

During program design, we sent the patterns with the detailed previous actions and raw events to the LLM so it understands the context in which this pattern should occur, and can generate a program design that is more grounded in real user behavior.
We set an occurrence threshold to be 5, so that only the frequent patterns will be displayed as program candidates.
We used a temperature of 0.2 to allow more creativity in the design while still maintaining relatively stable and accurate results.

To optimize the user experience, we run program design and ranking every 5 minutes, right after pattern identification, ensuring users get the designs as soon as possible.

In program generation, \tool generates script supported by Tampermonkey and will be installed by the user after manually review and confirmation. Each generated program declares explicit @grant scoping for its capabilities.

\subsubsection{Limitations}
Due to the capabilities of Chrome browser extensions, the generated programs can only run within the scope of the page. Therefore, any pattern that requires interaction with the Chrome browser itself, such as changing system settings, is not supported. 
Another limitation is that the complexity of the generated programs is restricted. For a single program, the sequence of actions needs to be well-defined, so any pattern that requires complex UI interaction, for example, drawing a graph on a canvas, is not supported. We consider these to be less automatable and less common, so we focus on patterns that can be automated with the capability of a program that can make people's lives easier.

\section{Evaluation Rationale, Design, and Context}
We conducted a lab user study to evaluate the usefulness, effectiveness, and usability of \tool in real-world tasks performed by diverse users, and to determine whether the design goals can be met.
\tool is designed to be highly personalized, supporting diverse usage scenarios.
As such, we took a primarily exploratory and qualitative approach.
Overall, we wanted to answer the following research questions:
\begin{itemize}
    \item \textbf{RQ1:} What are the difference between automatable patterns recognized by \tool and users? (DG1)
    \item \textbf{RQ2:} How does \tool support the generation, review, and refinement of program design from discovered patterns? (DG2, DG3)
    \item \textbf{RQ3:} What kinds of programs can \tool generate successfully from program design? (DG4)
    \item \textbf{RQ4:} How do users experience and perceive \tool in comparison with vibe coding? (DG3, DG4)
    \item \textbf{RQ5:} How do users incorporate \tool-generated programs into their daily workflows over time? (DG4)
\end{itemize}

\subsection{Study Design}
\label{subsec:studydesign}

The study included three parts: passive data collection, a lab session, and an optional longitudinal follow-up, all conducted in compliance with our institution’s IRB requirements.

\mysec{Part 1: Passive Data Collection}
For the first part of the study, participants met with a researcher to provide consent for the entire study, then installed the \tool browser extension during a brief meeting.
The participants were asked to use their Chrome browser as usual, with \tool passively recording their browsing activity and capturing their usage patterns.
All raw data was stored locally on the participant's device.
The participants were also informed about privacy concerns and how to block recordings by pausing or using a block list of URLs.
Once they have enough data (12 patterns), usually after 3-7 days of usage, the participants were asked to contact the research team to schedule Part 2. They were asked to continue using the browser until the Part 2 session. 
The extension displayed a readiness indicator to signal the participants, but other features (1-4 in \Cref{fig:interface}) were not displayed to the participants.

\mysec{Part 2: Lab session}
Participants then attended a one-hour virtual lab session with a researcher. Participants were asked to record their screen and audio during this session. 

\vspace{-0.2cm} \noindent 
\paragraph{a) Reflection (5 mins)}
First, participants reflected on their routine browser workflows and identified tasks they believed could be automated. The researcher asked guided questions to help them refine and improve their ideas into more clearly defined, potentially automatable tasks.
The resulting list was later used to confirm the initiation gap and to evaluate the patterns identified by \tool.
\vspace{-0.2cm} \noindent 
\paragraph{b) Vibe coding (10 mins)}
Then, participants spent 10 minutes attempting to build a program to automate their chosen regular routine, using an AI-powered tool of their choice to help them maximize AI capabilities rather than overcome learning barriers when learning new coding tools.
We asked participants to think aloud and did not provide any help other than when they were at risk of making serious, irreversible errors. Participants were allowed to give up in the middle. If participants were actively exploring or trying one solution at the time limits, we allowed them to spend an additional two minutes. 
\vspace{-0.2cm} \noindent 
\paragraph{c) \tool walkthrough (5 mins)}
Then, participants were introduced to the \tool interface and reviewed the programs it had discovered. We presented a Hello World program and let participants interact with it to understand the interface and the process of reviewing and generating programs. 
\vspace{-0.2cm} \noindent 
\paragraph{d) \tool review (30 mins)}
Participants were asked to review and try five programs they would like to use, and assess the accuracy, usefulness through a survey form implemented as part of the \tool interface.
They were allowed to generate the program multiple times after editing the program description.
We also asked the participants to think aloud.
We did not provide any help during this process, except when participants are at risk of making serious, irreversible errors. 
We provided guidance if participants are unsure how to proceed within the interface. 
After each review, participants were asked to either share or remove the program and complete a questionnaire about the identified pattern and the program’s usefulness. Throughout the process, we encouraged participants to refine the program description and regenerate the design if the initial attempt is unsuccessful. 
To evaluate the usefulness and accuracy of the program generation process, participants were given sufficient time to explore and understand each program, while we kept a soft cutoff of five minutes per program to ensure balanced time allocation across all programs. 
\vspace{-0.2cm} \noindent 
\paragraph{e) post study survey and semi-structured interview (12 mins)}
Finally, we had a post-study survey with parallel 5-point Likert scale items for both vibe coding and the \tool, covering dimensions such as ease of use, output accuracy, confidence, and usefulness.
We ended the session with a semi-structured interview to further understand their thoughts on automation blindness, the specification process, trust, and privacy. 
Participants kept the Tampermonkey scripts if they wanted, and all other programs and the \tool extension were fully removed.

\mysec{Part 3: Longitudinal Follow-up}
Before the end of Part 2, we asked participants if they are interested in participating in the optional longitudinal study.
If they agreed, they received a follow-up survey on continued script usage, real-world usefulness, and issues encountered.

\subsection{Recruitment}

We recruited ten participants\footnote{Two participants dropped out midway in data collection due to schedule changes.}, through social media and university forums.
All participants were at least 18 years old, used Google Chrome daily, and were proficient in English.
Prior programming and AI tool experience were collected during recruitment, and we balanced participants by programming experience level, slightly favoring those with no experience. 
Participants were compensated with \$45 gift cards after completing Parts 1 and 2, and an additional \$5 for the optional Part 3 survey. If parts 2 ran longer than an hour, we paid them an additional \$12.5 per 30 minutes.

\begin{table*}[htbp]
 \small
  \centering
  \caption{All participants' demographics and programs generated.}
    \vspace{-\baselineskip}
  \label{table:Participants}
    \begin{tabularx}{\linewidth}{l p{1.4cm} p{1.4cm} p{1.5cm} r r r X}
    \toprule
    ID & Program & AI tool & Major & Days & \# Act. & \# Pat. & Worked Programs \\
    \midrule
    P1 & Intermediate & Advanced & SWE &5& 2656 & 24 & googleAiOverviewExpander, youtubeAutoFullscreen \\
    \addlinespace
    P3 & None & Intermediate & Pharm. Sci. &5& 2450 & 13 & forwardGmailEnrollmentProof, automateTubiPlayback, automateYouTubeEngagement \\
    \addlinespace
    P4 & Intermediate & Advanced & SWE &5& 5666 & 30 & overleafAutoRecompile, acmDlAutoOpenPdf, arxivAutoOpenPdf \\
    \addlinespace
    P5 & Beginner & Beginner & Bus. Econ., Psychology &6& 2583 & 21 & facebookMarketplaceAutoInsights, facebookMarketplaceAutoRenewer, americaLearnsAutoTimesheet, whentoworkKeyboardNav \\
    \addlinespace
    P6 & None & Intermediate & Bio. Sci. &6& 3293 & 19 & googleSearchTabNavigator, stembleQuizNavigator, googleSearchAIExpander, wikipediaBirdRedirect \\
    \addlinespace
    P7 & Intermediate & Advanced & Biostatistics &4& 1595 & 21 & chatGptResponseInteraction, chatGptProjectNavigator \\
    \addlinespace
    P8 & None & Beginner & Law &5& 1117 & 12 & googleNewsNavigateCategory, zotGptAutoSendMessage \\
    \addlinespace
    P9 & Expert & Expert & SWE &8& 3397 & 35 & arxivAutoOpenPdf, googleScholarAutoSortYear, githubCopyCloneUrl, googleSearchExpandAIOverview \\
    \bottomrule
    \end{tabularx}
\end{table*}

All participants were all students (4 undergraduates, 5 graduates). P6 and P8 used Windows, while the rest used macOS. The remaining participants had a good mix of programming and AI tool experience. Detailed demographics and the programs generated by each participant can be found in \Cref{table:Participants}.

\subsection{Threats to Validity}

The vibe coding task was limited to 10 minutes, which may not have given participants sufficient time to fully iterate on their prompts and produce a polished result. 
However, our focus was on the specification process, rather than the final quality of the vibe-coded artifact. 
Capturing a complete vibe coding session would require substantially more time than is feasible in a controlled lab setting.

\tool's data collection period spanned approximately 4-8 days of active browsing. 
While this was sufficient to surface frequently recurring patterns, it may not capture important but infrequent workflows such as monthly reports or seasonal tasks, and likely overrepresents short-term daily routines. 
This constraint was driven by practical considerations for study scheduling. 
We note that the limited collection period likely underestimates \tool's potential: with longer logging and continuous use, the system would accumulate more diverse patterns and produce a richer set of program recommendations.

We allowed participants to use the vibe coding tool of their choice, introducing variability across participants. 
This was a deliberate decision: the learning curve for these tools is steep, and given the 10-minute time constraint, requiring all participants to use an unfamiliar tool would conflate tool learnability with the specification process we aimed to study. 
Letting participants use tools they were already comfortable with provided a more realistic picture of how they express their needs.

During the \tool review phase, participants selected at least five programs they were most interested in from the full list of discovered programs, rather than reviewing all of them exhaustively. 
While this is based on the practical considerations to keep the lab session length to be reasonable, this reflects realistic usage behavior.
In practice, users engage selectively with system recommendations rather than reviewing every result, much as they do with search engine results in information retrieval.

Even though we designed some privacy protection as descibed at \cref{subsec:privacy}, we acknowledge that the system still has residual privacy and security risks, including limited PII redaction, the potential for behavioral profiling or workplace surveillance misuse, prompt injection, malicious or buggy script generation despite scoped prompts, and risks from local compromise.  Future work could further explore security-related warning interface design and PII data reduction under this scenario. Meanwhile, any future deployment of Motif would require additional operational safeguards, such as organizational policies, audit logging, and tighter grant sandboxing.

The security scope of the generated programs is determined by the LLM and manually reviewed by humans. Although our prompt is designed to avoid generating programs that perform actions beyond program design, it remains possible that a generated program may perform harmful operations. In future work, a non-LLM-based deterministic mechanism could be implemented to determine and validate the program's security scope before deployment.

To ensure Motif is widely available with all participants, we used a cloud-based LLM (Gemini) for the study. However, in practice, most LLM calls made by our program, such as action abstraction, are small, typically invoking fewer than 3K tokens per request.
Based on our evaluation using a local LLM (Qwen3.6-35B-A3B), we can achieve performance comparable to that of cloud-based LLMs. This makes local deployment a viable solution for addressing some privacy concerns also.

\section{Study Findings}
We analyzed both quantitative and qualitative data to answer the research questions.
For qualitative data, including the study transcripts and the programs generated, one author performed initial open coding on the transcripts, and the codes were refined through discussion with another author to derive key themes.

\subsection{RQ1: Pattern Identification}
\begin{figure}
\includegraphics[width=\linewidth]{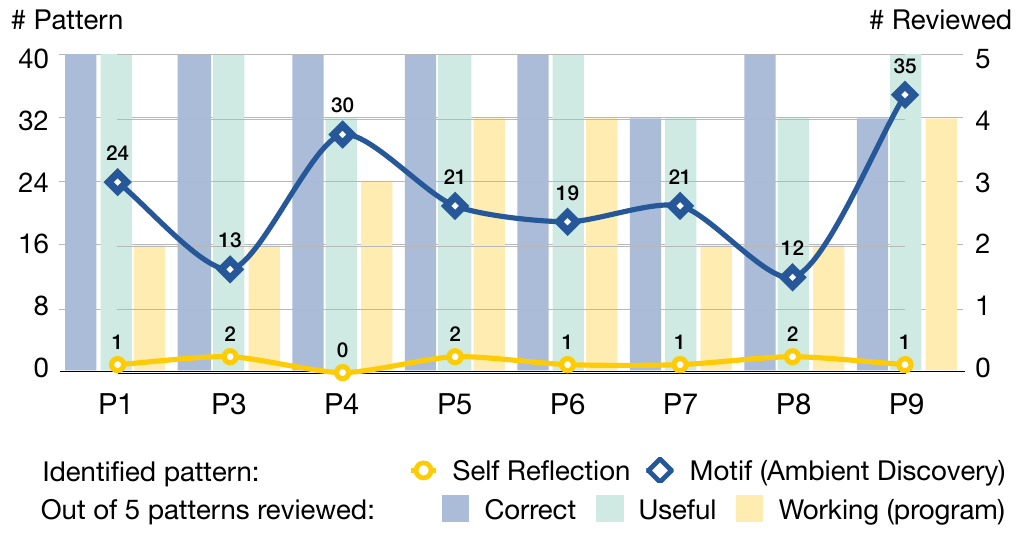}
\caption{Motif identified more automatable patterns (avg 22) than users' self-reflection (avg 1.3). Participants agreed that most of the patterns identified by Motif matched their routine (avg 4.75 out of 5) and are useful for automation (avg 4.6 out of 5).}
\label{fig:survey}
\end{figure}

\Cref{fig:survey} summarizes the number of patterns that participants could come up with during the self-reflection session, and that the \tool passively identified from the web usage logs.

At the beginning of the Part 2 lab session, participants were asked to self-reflect on their routine browser workflows and identify tasks they believed could be automated. 
All participants could come up with at least one routine to automate with only P4 explicitly saying no idea initially. 
However, except for P7, every participant originally only came up with a very broad idea, such as \textit{``browse a series of pages''} (P3), that cannot be automated with any tools available.
The researcher then helped participants develop ideas by asking their routine in more details, providing in-context examples, or asking guiding questions, such as "Which pages do you visit?" and "What actions do you perform frequently?" The researcher also provided in-scenario examples to help clarify. Finally, three participants developed two ideas each and five participants developed one idea each.

\tool, on average, could identify 22 usage patterns that occurred more than 5 times during Part 1 recording, that are automatable.
Of the 5 patterns each participant reviewed in detail, participants agreed with an average of 4.75 pattern matches or partial matches their daily routine, and found an average of 4.6 patterns to be potentially useful if the program works perfectly. 

This confirms the initiation gap, especially the task habituation and the mental model gap, and shows that the users often do not realize what can be automatable.
    
\subsection{RQ2: Pattern Usefulness}
\mysec{Pattern Correctness}
We evaluated the correctness of the identified patterns based on the participants' responses to the questionnaire when they shared or removed the reviewed patterns.
All participants agreed that at least four patterns they reviewed matched their actual routine and that creating a program for these would be useful. Five participants agreed that all five patterns they tried matched their routine. Two participants identified one pattern that matched their routine but deemed it not useful for automation, even if the program worked perfectly. Result as shown in \cref{fig:survey}. 

Most participants appreciated the patterns identified through self-reflection that were not recognizable to them.
P6 suggests,\textit{ ``I think it's surprising how I could track what you normally browse and come up with shortcuts for those. I think that's pretty cool''}. 
Similarly, P4 also suggests \textit{``I found that the pattern identification in motif is very useful, because it's very subtle, so I didn't realize it, but if I can have a program to help me do that, I feel it's very useful.''}

\mysec{Longer-term usage}
During the interview, six out of eight participants indicated that they want to continue using \tool, and they are interested in seeing more patterns that the system can discover.
Some participants (P4 and P6) indicated that a longer observation would be more useful to discover more patterns. \textit{``It will be interesting to have a leaderboard [so] I can know what I've done last month''} (p4)
However, participants also pointed out privacy concerns for the longer-term usage; \textit{``then there's other things, like security-type concerns''} (p8).

\mysec{Gap between reflected patterns and \tool detected patterns}
Interestingly, most of the patterns that participants reflected on during the reflection phase were not detected by \tool, and most of the patterns that \tool detected were not reflected by participants. Diving deeper, the differences mostly stem from the scope of the patterns. Our \tool focuses more on automatable patterns that are well-scoped, predictable, and shorter in sequence, while participants reflected on patterns that are larger in scope, more open-ended, and sometimes all-in-one solutions. For example, P7 reflected on a tool that automatically generates a job summary, performs translation based on the job description, and evaluates her own fitness for the position. In contrast, \tool detected several job refinement-related patterns. P7 suggested \textit{``[Motif] break it down into very, small steps, small patterns''}. 

\mysec{Intention $\ne$ Execution}
Users’ intended requirements were sometimes much more abstract than the actions they actually performed. For example, P8 initially identified a pattern regarding "ease of access" across different result categories on Google News. However, while engaging with the vibe coding prompt, she reflected and narrowed this abstract idea into a conceptual need: an auto-updating news summary. Even though \tool successfully generated a program based on her original recorded actions, P8 deemed the result unhelpful because it failed to reach this broader conceptual idea, noting: \textit{``It saves time jumping between pages, but the top bar is present with the floating bar.''}

\mysec{Hesitate of Modification} 
\tool was designed to provide precise automation while leaving room for users to edit; however, participants generally preferred regenerating programs rather than modifying textual descriptions to match their needs. For example, P7 tested a button-invoked program designed to transform Gemini into a "career station" by appending tailored prompts to copied job descriptions. During the trial, she noted the program's narrow scope, stating, \textit{``In this description, it seems like it just tells Gemini to be the career coach, not that if I switch to any AI agent, it will automatically tell them to be my career coach for me.''} Although P7 could have modified the trigger condition to generate a more generalized program that matched her broader intention, she did not do so. Overall, four participants modified the descriptions of just eight programs in total.

\mysec{Early Feasibility Judgement}
We found that some users prefer to assess the automation's feasibility based on the description.
By assuming that the functionality would not work without trying or modifying any program description, P8 rejected a pattern that could automatically highlight text on some websites, presuming that \textit{``[The program] need to understand the text in order for the highlights to work.''}

\mysec{Sorting by Usefulness Score is not enough}
Despite presenting patterns ranked by usefulness scores from high to low, participants did not simply select the top-ranked programs. Instead, they browsed the entire list and selected programs based on personal interest, with final selections likely distributed normally across the entire list. 
P4 even asked, `\textit{`Do you have some sorting for the available programs?''} 
Upon knowing the list was ranked by usefulness, P4 expressed surprise at patterns \tool identified as most useful but had not recognized in self-reflection. 
This suggests that while ranking provides structure, participants rely on personal judgment and curiosity rather than algorithmic scores when selecting which automations to explore. 
It may also suggest a frequency and usefulness combined sorting is not enough for recommendation the best patterns to users.

\subsection{RQ3: Program Usefulness}

Out of five programs each participant reviewed, three worked as expected on average. 
Among the 40 programs participants reviewed, the LLM labeled 1 as high difficulty, 22 as medium difficulty, and 17 as low difficulty. Low difficulty programs are more likely to work as expected, with 82\% working as expected, compared with only 45\% for medium difficulty programs. 
The only high difficulty program did not work as expected.

\subsubsection{Programs that function as expected}

From the entire workflow, locating and designing a correct trigger is critical for the overall success of the generated program. 
Among the 40 programs generated by participants, 30 were designed to use a button as the trigger, seven were designed to run automatically when reaching a certain page, and five were designed to be triggered by keyboard shortcuts.
Among these, on page load and shortcut triggers showed higher success rates: 86\% of on page load-triggered programs worked as expected, and 67\% of shortcut-triggered programs worked as expected, while only 54\% of button-triggered programs worked as expected. This may be due to the fact that most existing Tampermonkey scripts used for LLM training were designed for full automation rather than user interaction.

Since the built programs were personalized for each user based on their browsing history, the accepted programs were highly diverse by type. 
Some accepted programs were for routine work that people needed to perform frequently, like automatically filling out the timesheet with a template and submitting that (americaLearnsAutoTimesheet). 
Or highly repetitive work that was personalized to each user, such as automatically opening a PDF page whenever they reach the ACM Digital Library (acmDlAutoOpenPdf), or automatically sorting by years whenever they open Google Scholar (googleScholarAutoSortYear). Some interaction with a video streaming website, such as automatically selecting the next episodes and automatically playing the video (automateTubiPlayback), is also highly accepted. 
A last but not least type is a shortcut program that helps people jump to the right place faster, such as automatically navigating between different types of Google search results with a shortcut (googleSearchTabNavigator).

\subsubsection{Programs that did not work}
Only one participant removed a program because it did not work. This program was a script designed to join Zoom meetings without user confirmation. 
It failed because the task requires interacting with the Chrome browser itself by interacting with a browser pop-up window, which is outside the scope of a Tampermonkey script, since it only runs in the isolated context of a specific webpage.

Among all programs, we did not observe any program that will perform a non-irreversible action without user confirmation, such as sending emails automatically without the user clicking send.
We believe the program design prompts are in effect to prevent those security vulnerable programs from being generated.

Some websites have complex or dynamically loaded DOM structures that the original website intentionally designed to prevent automation. For example, P8 evaluated a pattern called calendlyScheduleEvent, which is designed to automatically fill out the scheduling form on Calendly and submit it, but it did not work because the form is dynamically loaded, and the program cannot locate the right element to fill out and submit.

Sometimes, the pattern users selected are too complex for the program to work; for example, P7 selected geminiWorkflowAutomation, which requires switching between Gemini models (Fast/Thinking) and streamlines the process of updating previous prompts by feedback and automatically submit the new version. P7 tried to regenerate the design several times, did not attempt to correct the program description, and could not make the program work as expected. We think this is because the program is too complex, making it hard for the LLM to capture the real user intent when designing and generating the program. 

\subsubsection{Perceived Usefulness}
Overall, five participants (P3, P4, P5, P6, P9) had positive experiences and generated three to four programs that are very useful. 
P3 suggested that, \textit{``With Motif, I thought that it was pretty easy (to generate the program) ... 
I was surprised that they would think some of them could be automated, like, [describing the forwardGmailEnrollmentProof pattern].''} 

Three participants (P1, P7, P8) were able to generate only two not very useful programs and had negative experience with the entire process. 
We think this is due to participants' usage patterns. 
Two participants had concerns about privacy and did not use \tool for all of their work; they created multiple Chrome profiles and only used the one with \tool for some specific tasks to prevent privacy issues. 
One of these two was also concerned about privacy issues of trying some patterns during the study in front of us and chose to try rather simple programs with limited utility, such as automatically sending a chatbot message. Even for people who stand strongly for privacy concerns, they are still willing to share the generated program with others, which confirms our design goal of using programs to maintain privacy. Another participant did not use English in their daily workflow, given the majority focusing of Gemini and our workflow is on English content, the entire workflow may be less accurate when the user is not using English, which may cause the entire Motif workflow doesn't run as expected.

\mysec{Potential time saving and personalization}
Among the generated programs, participants appreciate the time it could potentially save, with P3 suggesting, \textit{``It took me a while to draft the forwarding email to my employers, so it's nice to have it automated as I need to do this every quarter for my job.''} Sometimes, the program can personalize and remember the last step of the user and continue from there, such as P3 suggests, \textit{``This is useful because it remembers what episode I am at in the series.''}

\mysec{Too small action may not be beneficial}
If the action the program performs was too small, the participant complains about the actual effort saved by it, such as P8 suggested ``[the program] duplicates the send button. They are in the same location.''

\mysec{Beginners, no experience, and experts benefit more than intermediate experience people}
For people with less programming experience, they appreciate the ease of use, and specifically thanks to the regeneration feature, which allows them to easily modify the program description without going through the entire description. P3 has no experience in programming and suggested \textit{``I also like the regenerate design, how I can just press it, and then it would try to fix it, so I don't really have to do anything too difficult.''} Beginners and no experience people also appreciated the tutorial with program design, which tells them how to use the program, as P5 suggests \textit{``it easy for people to use and understand, because I don't really have a background in code, but, like, having that tutorial thing and, like, having you explain how it works, I think it was really helpful.''}

Expert care more aspects about the program being generated than other people, such as the performance, and appreciate the user under control design and easy to modify, test and try of \tool. P9 is an expert programmer who has the same pattern from self-reflection and \tool, he recognized that \textit{``\tool's program is faster than [the vibe coding solution] I installed. It's very fast.''} He also recognizes the ease of modification, saying "Directly edit the program by natural language is easy to use... [\tool] is easy to correct the program."

However, people with intermediate experience are more likely to be critical of the program generation process and wish to stick with their original solution. 
P1 suggested ``I think programming in natural language… but usually I would ask another… AI to, like, generate the exact prompt of the function I want". They also wish to have natural language to modify the program description, as P7 suggested \textit{``If you can add an AI toolset, help me translate my natural language into a more specific prompt to the program description, it will be easier.''  }
They also complain about the waiting time of program generation, as P1 suggested, \textit{``Sometimes the waiting time is different from each trial, so it kind of makes me confused, like, is it working or not?''}

\subsection{RQ4: Comparison with Vibe Coding}

\begin{figure}
\vspace{-\baselineskip}
\includegraphics[width=0.9\linewidth]{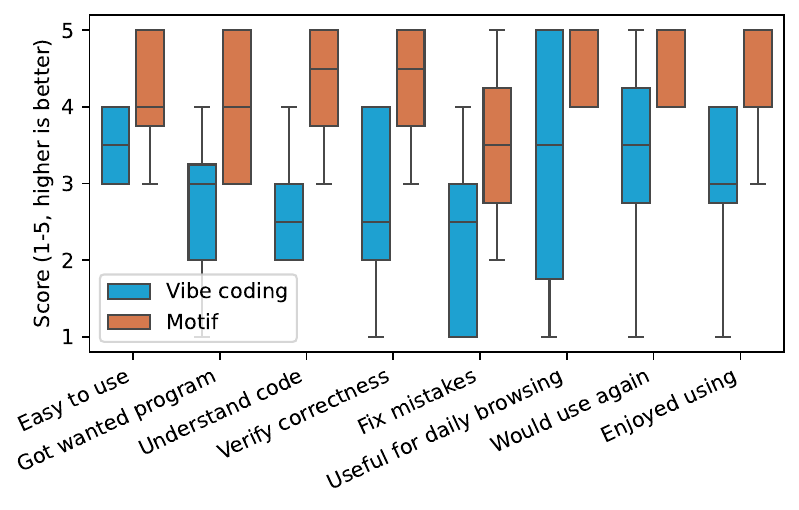}
\caption{Post-study survey results. Most participants reported higher satisfaction with \tool compared to vibe coding across all dimensions.}
\vspace{-\baselineskip}
\label{fig:post-study-survey}
\end{figure}

The overall subjective user feedback is shown in \cref{fig:post-study-survey}. Overall, participants reported high satisfaction with \tool's usability and effectiveness in generating automation programs for their routine web tasks, compared with vibe coding. \tool received higher ratings than vibe coding across all dimensions, with larger differences in understanding the code, verifying correctness, and enjoyment. This aligns with interview; for example, P4 said, ``I think the title and the description are good, because it's very specific, and it just describes a very, very specific behavior, not very general, so I can understand.'' People generally enjoyed using \tool, which was further confirmed in the subsequent interview where six out of eight participants wished to continue using \tool, and the others expressed interest in using it if the privacy concerns or program accuracy could be improved.

\mysec{Vibe coding vs. \tool}

Comparing vibe coding and \tool in general, most users with no experience or beginners found vibe coding difficult and prefer to use \tool. P6 was able to get a working program and suggested that ``I would rather not do the ChatGPT one. Cause I like how this one just… Generate stuff for you, and you don't have to manually type it in.'' Fortunately, \tool also detected the same pattern for her. When comparing the entire process of both approaches in terms of trustworthiness, she suggested ``probably [\tool], because ChatGPT made, like, 3 mistakes.''

Surprisingly, only two out of eight people were able to use the vibe coding tool they selected to build the workflow successfully. Participants used a variety of vibe coding tools, like ChatGPT, codex, which recommended a wide range of solutions, such as Playwright (P4), the Claude Code in-app widget (P1, P7), a third-party no-code platform (P3), Chrome extension (P9) and Tampermonkey (P6). P6 and P9 were able to successfully set up the environment following LLM guidance and finish all desired functionality, most likely due to the correct selection of the tool should be used (e.g., Tampermonkey is more suitable for web automation in user's personal browser). Four participants had setup issues that prevented them from successfully using the vibe coding tool, with two of them (P1, P7) experiencing issues with the Claude Code widget, likely due to incorrect or incomplete Claude Code setup. The other two participants (P3, P5) had issues with setting up the recommended platform or toolset on their local machine, which may be due to the complexity of the setup process or lack understanding of computer architecture. Although P6 successfully setup the environment, she still had to debug the setup 2 times and ask for guidance, which suggests that the setup process is still not smooth and straightforward in the vibe coding environment. 

Unlike providing a fixed pattern description in \tool, the vibe coding process is more iterative and open-ended, offering a significant benefit. Participants (P1, P4, P7) utilize the iterative chat feature to refine and formulate more fine-grained ideas after seeing the system's initial responses, establishing a collaborative dynamic between themselves and the vibe coding platform. Users also appreciate the high flexibility of vibe coding, which allows them to express their needs in a more open-ended manner and receive more comprehensive solutions. For example, P7 suggested, \textit{``If the vibe coding one did work, I might prefer that one, because that looks like a really big and comprehensive app that I can use''} and P4 suggest "[vibe coding tool] is more straightforward for me to interact with".

Besides the setup challenging, when comparing the core programs successfully generated by both approaches, the vibe coding solutions are significantly shorter in terms of lines of code, but they are also less informative and provide less feedback to the end-user. For example, the vibe coding solution for P6 is a 3-line Tampermonkey script that automatically redirects from one page to another without any user confirmation or notification. In contrast, the program generated by \tool is much longer and provides an interactive entry point for the user to initiate the transition. The extreme automation by vibe coding could pose a malicious program by essentially preventing the user from accessing the original page.

\subsection{RQ5: Long-term usefulness}
We sent a follow-up survey 2–4 days after participants completed the study session, asking about their continued use of the worked programs generated by the \tool. Six participants (P1, P3, P4, P5, P8, P9) responded to the follow-up survey. Among 18 programs they had kept and used, they regularly used 30\% of programs, used 40\% of programs a few times, and had not yet had the opportunity to use the remaining programs. Regarding reasons for non-use, two participants (P3, P8) reported four programs to be less useful than expected, and P5 reported not having had the opportunity to use two programs.

\begin{figure}
\includegraphics[width=0.9\linewidth]{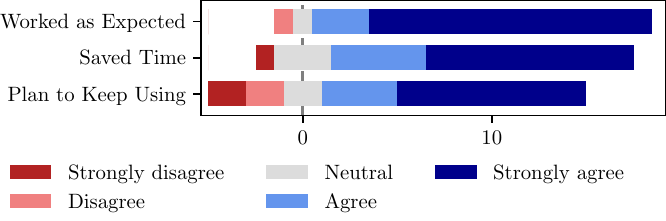}
\caption{Followup survey results.}
\label{fig:followup}
\end{figure}

We asked participants, using a Likert scale, whether the program continued to work as expected, saved time, and whether they would continue using it. The results are shown in \cref{fig:followup}. Overall, most of the programs that participants kept continued to work as expected, saved time, and participants were willing to continue using them. By analyzing the programs that participants were less likely to continue using and that were deemed not to save time, such as zotGptAutoSendMessage, we observed that participants were less likely to continue using programs that performed only minor actions or interfered with the original functionality. Another pattern among programs that participants were less likely to continue using is those that have too many actions at once, such as automateYouTubeEngagement, which automatically likes, subscribes, and shares content in one click. These behaviors can be overwhelming for users and may even cause technical issues.

One of the most far-reaching influences of this experiment is that it might reshape how people think about automation and patterns. In the follow-up survey, all participants agreed that using \tool changed how they think about their browser workflow, with four participants strongly agreeing. Two participants also came up with new tasks that they could automate after using \tool. However, upon reviewing the new patterns that participants self-reflected on, only one of them was a valid, automatable pattern, suggesting that people's thinking about their workflow is still limited by the mental model gap and task habituation, even after using \tool. 

\section{Discussion and Implications}

\mysec{Mixed method in automation} 
One drawback of the program generation is its lack of compatibility with complex DOM websites, such as Google Docs, games, or canvas drawing pages. 
P1 tried to write a program to automatically download a Google Doc as a PDF, but it did not work because the program could not locate the element. 
One possible solution is to combine the program with a computer-use agent to recognize and perform clicks in real time, which can significantly increase the program's compatibility while still maintaining its benefits.

\mysec{Higher level of abstraction and fine-grained program generation} 
From the gap between the reflected patterns and those detected by \tool, we can see that people tend to think in terms of high-level, open-ended, all-in-one solutions, while \tool focused on automatable patterns that are well-scoped, predictable, and shorter in sequence. 
Although users could modify the program design in our interface, they often seemed unwilling to do so, they preferred
to click ``regenerate'' 
The reasons are twofold: first, users with limited programming experience often found it difficult to understand the program description; second, users with intermediate experience perceived the cognitive load of modifying it as too high.
This highlights the need for a higher level of abstraction in program design, which could better capture user intent and generate programs more closely aligned with their needs.
Given that the goal of the current work was to test the feasibility and benefits of the new paradigm of ambient automation discovery, we plan to keep this as future work and explore it further.

\mysec{More Intuitive Program Refinement}
P1, P7, and P8 expressed a desire for a system that allows them to modify program descriptions using natural-language prompts, or a place where they can indicate what went wrong. 
As P8 suggested, ``I would say maybe just a place to input what's going wrong, because right now you only have whether this program works or it doesn't.'' This could be a very intuitive feature for users to quickly improve the generated programs.

\section{Conclusion}

In this paper, we presented \tool, a system that automatically identifies routine web usage patterns and generates personalized automation programs. Through a user study with eight participants, we found that \tool can identify many more automatable patterns than users can self-reflect on, and that most of the identified patterns were useful for automation. 
We also found that most generated programs worked as expected and were perceived as useful by participants in long term. 
Our findings suggest that \tool can effectively recognize and automate routine web usage patterns, thereby significantly enhancing user productivity and experience. 
Future work should explore ways to further abstract user intent, improve program generation, and streamline the program adoption workflow.

\bibliographystyle{ACM-Reference-Format}
\bibliography{daye}

\appendix

\section{Prompt used}

\begin{tcolorbox}[title=Abstracting actions]
You are an AI analyzing a user's web activity. 
Based on the provided UI events and the attached screenshots, summarize the step-by-step actions the user took in this time period.

NOTE: The 'path' in the 'screenshot\_taken' events corresponds to the URIs of the images attached to this message. Use the timestamps and sequence of events to map the actions to the correct images. The red dot inside the images stands for the cursor position at the time of the screenshot.

=== UI EVENTS ===
\$\{uiEventsString\}

=== CORRELATED NETWORK CONTEXT ===
\$\{networkContextString\}

Output your response as a concise list of sequential actions meets the requested JSON schema. For each action:
- "action": Provide a concise, human-readable description of what the user did.
- "meta": Extract relevant technical identifiers to help locate elements or replicate the action (e.g., target URL, exact text typed, element IDs, DOM classes, or correlated network payload keys).
\end{tcolorbox}

\begin{tcolorbox}[title=Identifying patterns]
Analyze raw browser logs and compile deterministic sequences into Tampermonkey scripts.

=== RULES ===
1. EXTRACT CODE: Write robust scripts (with // ==UserScript==) for automatable tasks (clicks, forms, APIs). Ignore random noise or visual reading. Summarize the code in plain English.
2. MAP TRACES (Array of Arrays): Chronologically group log 'id's that the script automates. Maintain temporal isolation. Example: If a task happens twice, output [["1", "2"], ["8", "9"]]. NEVER merge distinct executions.
3. MATCH STATE: If the sequence matches a KNOWN\ PATTERN, output its exact name in 'existing\_pattern\_match'. If it is new, output 'NEW\_PATTERN' and provide a camelCase 'pattern\_name'.

=== DATA ===
KNOWN PATTERNS:
    \$\{existingPatternSummaries\}
    
NEW BATCH TO ANALYZE:
    \$\{newLogs\}
\end{tcolorbox}

\begin{tcolorbox}[title=Program design]
You are an Expert Browser Automation Architect and UX Designer. 
I am providing you with multiple historical execution traces of a repetitive user workflow ("\$\{pattern.name\}").

Your task is to analyze these executions ('all\_historical\_executions') to understand the core user intent. Identify the common denominator and abstract away any noise, misclicks, or highly specific one-off values (e.g., dynamic\ IDs, specific search strings).

CRITICAL INSTRUCTIONS:
1. Optimize for User Experience (UX): Do not just blindly automate the exact sequence. Think about the most user-friendly way for someone to interact with this script. Instead of aggressive auto-execution on page load, consider better patterns like injecting a floating "Execute Task" button, using a keyboard shortcut, or waiting for a specific contextual UI element to appear before acting.
2. NO TECHNICAL JARGON: Describe the user journey in plain English (e.g., "clicks the search bar"). You are strictly forbidden from using HTML tags, CSS classes, XPaths, or DOM locators.
3. Concise Description: Provide a short, clear description of the logical workflow and how the user will trigger it. Focus entirely on the human-computer interaction.
4. Feedback Mechanism: Decide exactly when the script should trigger a message notification to inform the user of the script's state (e.g., completion, error, need attention).

Finally, evaluate how feasible this is to automate via Tampermonkey, keeping in mind out of scope, complex iframes, or SPAs (Single Page Applications).
\end{tcolorbox}

\begin{tcolorbox}[title=Pattern Ranking]
You are an Automation Architect. Evaluate the following \$\{totalPatterns\} patterns and predict their future automation ROI (1-10).

DATA CONTEXT:
- 'occurrences': A 2D array. Each inner array is a single execution. The timestamps inside represent individual steps taken during that execution.

CRITERIA for your score and reasoning:
1. Volume \& Pace: High frequency, dense clusters, and rapid steps indicate high ROI.
2. Predictability: Ongoing, continuous needs score higher than one-off transient events.
3. Likelihood of happening in the future: Patterns that are fundamental to user workflows or show signs of increasing frequency are more likely to be automated and thus score higher.

Provide an automation\_value\_score (1-10) and a brief reasoning addressing the 3 criteria above.

There are exactly \$\{totalPatterns\} patterns in the data below. You MUST evaluate every single one. Do not skip, summarize, or omit any patterns. Your final JSON array must contain exactly \$\{totalPatterns\} objects.

Data:
\$\{dataForLLM\}
\end{tcolorbox}

\begin{tcolorbox}[title=Program Generation]
You are an Expert Tampermonkey Developer. Write a robust userscript based on the provided BLUEPRINT and TRACES.

=== REQUIREMENTS ===
1. Behavior \& UX: 
The BLUEPRINT 'program\_description' dictates the final user experience and goal. You MUST adhere to the intended UX (how it triggers and notifies). For 'notification' situations, prefer Tampermonkey's `GM\_notification` (include `// @grant GM\_notification`).

2. Execution Toolkit: 
You decide the best technical approach to achieve the program\_description goal based on the provided traces. Choose the most reliable and efficient strategy from the following viable options:
    - Network Requests: Consolidate manual clicks into background \`fetch()\` or \`GM\_xmlhttpRequest\` calls. Safeguard: Only simulate state-changing requests (POST, PUT) if the exact endpoint and payload are captured in the traces. Do not guess payloads.
    - Shortcuts: Leverage official keyboard shortcuts if they are faster/safer, invoking them via keyboard events.
    - DOM Targeting \& Manipulation: When locating UI elements is required, you MUST use an **Anchor-First (Bottom-Up)** approach. Locate a highly stable element first, then traverse relatively if needed. 
        - Identify multiple high-quality candidate anchors from the TRACES (e.g., Content-Based like aria-labels/visible text via Xpath, and Property-Based like semantic classes/IDs). All selectors MUST be semantically meaningful.
        - Then evaluate these candidates and explicitly select the SINGLE best, most robust selector to implement. Prefer content based selector. Any ID, class, or text used MUST be explicitly captured and proven from the TRACES.
    *Note: ALWAYS wrap DOM selections and interactions in async wait functions.*

3. Planning First: 
BEFORE writing any code, you MUST output your thought process using two strict XML blocks: <plan> and <automation\_steps>.
    - Inside <plan>: Analyze the traces and the scenario, evaluate the trade-offs of the strategies in your toolkit, and explicitly state which approach is best for this specific task and why.
    - Inside <automation\_steps>: Map out the technical sequence for all scenarios based strictly on the TRACES. You should also plan for SPA rendering delays. For each step, use this exact pipe-separated format:
      [Action Name] | [Operation: Detail the exact execution strategy. If DOM, list final selected anchor. If Network, list the exact endpoint/payload. If Shortcut, list the exact keys. If the previous action triggered a UI change or network request, this step MUST explicitly define a Wait/Sleep operation.]

4. Format Requirements: 
You MUST output in exactly two parts.
    PART 1: The <plan> block, followed by the <automation\_steps> block.
    PART 2: Immediately following the </automation\_steps> tag, output ONLY the raw JavaScript code starting with the standard // ==UserScript== header. Do not add any conversational text before or after the code.

=== BLUEPRINT (WHAT TO BUILD) ===
\$\{blueprint.trigger\_condition, blueprint.program\_description, blueprint.notification\}

=== TRACES (HOW THE USER DID IT) ===
\$\{allExecutions\}
\end{tcolorbox}

\begin{tcolorbox}[title=Tutorial Generation]
You are an expert Technical Writer. Your task is to write a clear, user-friendly Markdown tutorial for a newly generated Tampermonkey userscript called "\$\{patternName\}".

=== CONTEXT ===
BLUEPRINT (What the script was designed to do):
\$\{blueprint\}

GENERATED SCRIPT (The actual code):
```javascript
\$\{generatedScript\}
```

=== REQUIREMENTS ===
Write a concise, easy-to-read Markdown document that includes:
1. **Overview:** A brief description of what the script does.
2. **How to Trigger It:** Step-by-step instructions on how the user activates the script (e.g., clicking a specific button, pressing a hotkey, or if it runs automatically on page load).
3. **What to Expect:** What the UI or page will look like after the script successfully executes.

Keep the tone helpful and accessible for a standard web user. Be concise and only provide the Markdown tutorial.
\end{tcolorbox}

\end{document}